\documentclass[twocolumn,prd,groupedaddress,nofootinbib]{revtex4}
\begin{document}

\title{Comment to: Corrections to the fine structure constant in the
spacetime of a cosmic string from the generalized uncertainty principle}
\author{Geusa de A. Marques$^{*}$}
\affiliation{$^{}$ Departamento de F\'{\i}sica, Universidade Federal de Campina Grande\\
58109-790 Campina Grande, Pb, Brazil.}
\date{\today }

\begin{abstract}
In the paper [F. Nasseri, Phys. Lett. B 632 (2006) 151--154], F. Nasseri
supposed that the value of the angular momentum for the Bohr's atom in the
presence of the cosmic string is quantized in units of 
$\hbar$. Using this assumption it was obtained 
an incorrect expression for Bohr radius in this scenario.
In this comment I want to point out 
that this assumption is not correct 
and present a corrected expression for the Bohr radius in this
background.
\end{abstract}

\maketitle

In a recent paper F. Nasseri\cite{1} analyzed the fine structure constant in
the spacetime of a cosmic string from the generalized uncertainty principle.
The author claims that the Bohr's radius in this system increases by a
factor of order $\pi /4\times 10^{-6}$. Such a result was obtained by the
author considering that the stationary orbits in the spacetime of a cosmic
string have an integer  number of wavelengths in the interval from $0$ to $2\pi $%
. This is not correct once for stationary orbits we have

\begin{equation}
\oint \frac{dS}{\lambda }=n;\; \; \; \; n=1,2,3,...  \label{1}
\end{equation}%
where $S$ is the length of \ the orbit and $\lambda $ the wavelength. As $%
\ $the metric of a cosmic string is given by

\begin{equation}
ds^{2}=c^{2}dt^{2}-dz^{2}-d\rho ^{2}-\rho ^{2}d\varphi ^{\prime 2},  \label{2}
\end{equation}%
where $\varphi ^{\prime }=\left( 1-\frac{4G\mu }{c^{2}}\right) \varphi ,$
which implies that $\varphi ^{\prime }$ varies from 0 to $2\pi b$.
Therefore $dS=\rho d\varphi ^{\prime }$, leads to a correction in the
condition of quantization of the angular momentum. Thus, instead of $%
L_{n}=n\hbar $ as considered by the author in \cite{1}, we have $L_{n_{(b)}}=\frac{n}{b}%
\hbar $, where $b$ is the deficit angle and is given by $b=1-\frac{4G\mu }{%
c^{2}}$\cite{2}. With this consideration, the radius of the $n$th Bohr orbit
of the hydrogen atom in the presence of a cosmic string becomes%
\begin{equation}
\rho _{n}=\frac{4\pi \epsilon n^{2}\hbar ^{2}}{me^{2}\left( 1-\frac{\pi }{4}%
\frac{G\mu }{c^{2}}\right) \left( 1-4\frac{G\mu }{c^{2}}\right) ^{2}}. 
\label{3}
\end{equation}%
The eq.(17) of \cite{1} is correct only for flat spacetime. In the presence
of a cosmic string, the radius of the nth Bohr orbit of the hydrogen atom is
given by 
\begin{equation}
\hat{a}_{B}=\frac{4\pi \epsilon \hbar ^{2}}{me^{2}\left( 1-\frac{\pi }{4}%
\frac{G\mu }{c^{2}}\right) \left( 1-4\frac{G\mu }{c^{2}}\right) ^{2}}. 
\label{4}
\end{equation}%
In the absence of a cosmic string, the lowest orbit ($n=1$) of the Bohr
orbit has the following expression 
\begin{equation}
a_{B}=\frac{4\pi \epsilon \hbar ^{2}}{me^{2}}=5.29\times 10^{-11}m.  \label{5}
\end{equation}%
Thus combining eqs.(4) and (5), we obtain%
\begin{equation}
\frac{a_{B}}{\hat{a}_{B}}=\left( 1-\frac{\pi }{4}\frac{G\mu }{c^{2}}\right)
\left( 1-4\frac{G\mu }{c^{2}}\right) ^{2}.  \label{6}
\end{equation}%
In the weak field approximation, eq.(6) turns into 
\begin{equation}
\frac{a_{B}}{\hat{a}_{B}}\approx 1-\frac{G\mu }{c^{2}}\left( 8+\frac{\pi }{4}%
\right) .
\end{equation}

In the limit $\mu \rightarrow 0$, i.e., in the absence of a cosmic string, $%
\frac{a_{B}}{\hat{a}_{B}}\rightarrow 1$. It is worth noticing that there is
a difference between eq.(22) of \ref{1}  and eq.(6) of this Comment. Taking
into account the correction given by eq.(6) and inserting $\frac{G\mu }{c^{2}%
}\simeq 10^{-6}$ we get

\begin{equation}
\hat{a}_{B}=\frac{a_{B}}{\left( 1-\left( 8+\frac{\pi }{4}\right) \times
10^{-6}\right) }.  \label{7}
\end{equation}%

From the above equation, we conclude that the numerical factor which correct
the Bohr radius is different from the one obtained in  eq.(23) of
  reference \cite{1}.

\vskip 0.5 cm

{\large \bf{Acknowledgments:}} I would like to thank V. B. Bezerra for
the valuable discussions and to thank to CNPq-FAPESQ-PB (PRONEX)
 for partial  financial support.

\vskip1.0 cm

$^{\ast }$Electronic address: gmarques@df.ufcg.edu.br

\end{document}